%% file: main.tex
\documentclass[10pt, conference]{IEEEtran}
\IEEEoverridecommandlockouts
\usepackage{cite}
\usepackage{amsmath,amssymb,amsfonts}
\usepackage{algorithmic}
\usepackage{graphicx}
\usepackage{textcomp}
\usepackage{xcolor}
\usepackage{booktabs}
\usepackage{url}
\usepackage[many]{tcolorbox}
\usepackage{indentfirst}
\usepackage{float}
\usepackage{tabularx}
\definecolor{darkgreen}{rgb}{0.0, 0.5, 0.0}

\definecolor{stageone}{RGB}{20,90,210}
\definecolor{stagetwo}{RGB}{0,125,40}
\definecolor{stagethree}{RGB}{125,45,195}
\definecolor{stagefour}{RGB}{180,75,0}
\definecolor{stagefive}{RGB}{180,50,115}
\definecolor{stagesix}{RGB}{0,115,120}
\definecolor{stageseven}{RGB}{15,85,195}
\definecolor{stageeight}{RGB}{160,100,0}

\usepackage{cleveref}

\usepackage{tikz}
\usetikzlibrary{arrows.meta, positioning, shapes.geometric, fit}
\usepackage{adjustbox}

\newcommand{\toolname}{\textsc{Flow-A11y}}

\newcommand{\stagebadge}[2]{%
  \raisebox{0.02ex}{%
    \tikz{
      \node[
        circle,
        fill=#1,
        text=white,
        font=\bfseries\fontsize{7}{7}\selectfont,
        inner sep=0pt,
        minimum size=3.5mm
      ] {#2};
    }%
  }%
}
\newcommand{\stageone}{\stagebadge{stageone}{1}}
\newcommand{\stagetwo}{\stagebadge{stagetwo}{2}}
\newcommand{\stagethree}{\stagebadge{stagethree}{3}}
\newcommand{\stagefour}{\stagebadge{stagefour}{4}}
\newcommand{\stagefive}{\stagebadge{stagefive}{5}}
\newcommand{\stagesix}{\stagebadge{stagesix}{6}}
\newcommand{\stageseven}{\stagebadge{stageseven}{7}}
\newcommand{\stageeight}{\stagebadge{stageeight}{8}}

\newtcolorbox{rqresultbox}{
  colback=white,
  colframe=black,
  boxrule=0.45pt,
  arc=0pt,
  outer arc=0pt,
  left=4pt,
  right=4pt,
  top=4pt,
  bottom=4pt,
  boxsep=0pt,
  before skip=4pt,
  after skip=5pt,
  width=\columnwidth
}

\def\BibTeX{{\rm B\kern-.05em{\sc i\kern-.025em b}\kern-.08em
    T\kern-.1667em\lower.7ex\hbox{E}\kern-.125emX}}
\begin{document}

\title{\toolname{}: Flow-Aware Accessibility Testing}

\input{sections/authors}

\maketitle

\begin{abstract}

Modern web applications increasingly expose accessibility barriers through interaction flows rather than static page snapshots. Keyboard traps, focus loss, modal leakage, delayed status updates, dynamic controls, and changing page regions often become observable only after users perform concrete actions. These behaviors are directly related to dynamic WCAG criteria, yet they remain difficult to automate because their assessment depends on runtime interaction evidence and is still commonly performed through manual inspection.

We present \toolname{}, a flow-aware accessibility testing system for interaction-dependent WCAG criteria. Given a target page and a natural-language scenario, \toolname{} executes the flow in a real browser, records an ordered runtime trace, constructs criterion-specific evidence packets, gates unsupported judgments, and emits auditable findings grounded in resolvable runtime evidence. Evaluated on 19 real public-web scenarios covering 45 dynamic WCAG criteria, \toolname{} achieves over ten times higher oracle agreement than a generic browser-agent audit, while its evidence-calibration layer improves fail precision from 23.5\% to 41.4\% and eliminates invalid evidence references.

These results show that runtime traces provide actionable evidence for assessing interaction-dependent accessibility behavior. They demonstrate a practical path toward automating dynamic WCAG criteria that page-level scanners cannot assess and that have traditionally required manual evaluation.

\end{abstract}


\input{sections/introduction}
\input{sections/motivation}
\input{sections/background}

\input{sections/approach}

\input{sections/experimental_design}
\input{sections/results}
\input{sections/discussion}
\input{sections/threats}
\input{sections/artifact}
\input{sections/conclusion}

\bibliographystyle{ieeetr}
\bibliography{references}

\end{document}

%% file: sections/authors.tex
% \and
% \and
% \and
% \and
% \and
\author{
\IEEEauthorblockN{Nasr Eddine Fliti}
\IEEEauthorblockA{
\textit{University of Luxembourg}\\
Luxembourg, Luxembourg\\
nasreddine.fliti@uni.lu
}
\and
\IEEEauthorblockN{Leisan Kokorina}
\IEEEauthorblockA{
\textit{University of Luxembourg}\\
Luxembourg, Luxembourg\\
leisan.kokorina@uni.lu
}
\and
\IEEEauthorblockN{Florian Tambon}
\IEEEauthorblockA{
\textit{University of Luxembourg}\\
Luxembourg, Luxembourg\\
florian.tambon@uni.lu
}
\and
\IEEEauthorblockN{Mike Papadakis}
\IEEEauthorblockA{
\textit{University of Luxembourg}\\
Luxembourg, Luxembourg\\
michail.papadakis@uni.lu
}
}

%% file: sections/introduction.tex
\section{Introduction}
\label{sec:introduction}

Web applications have become the primary interface to essential services in domains such as e-commerce, e-government, education, and healthcare. As a result, web accessibility is no longer optional: it is a legal, societal, and engineering requirement affecting over 1.3 billion people worldwide living with disabilities\cite{report:who:disability-health:2023}. Accessibility obligations are codified in regulatory frameworks such as the Americans with Disabilities Act (ADA)~\cite{legal:us:ada-web-guidance}, Section 508 in the United States~\cite{legal:us:section508}, and the European Web Accessibility Directive~\cite{legal:eu:web-accessibility-directive:2016}, and are operationalized through the Web Content Accessibility Guidelines (WCAG)~\cite{standard:w3c:wcag21}. 
    
WCAG is widely accepted as the international normative reference for web accessibility and has been adopted, referenced, or mandated by legislation and procurement policies in many jurisdictions ~\cite{standard:w3c:wcag21,standard:w3c:wcag22,conf:issta:ma11y:2024}. The guidelines are organized around four foundational principles, often abbreviated as POUR: Perceivable, Operable, Understandable, and Robust~\cite{standard:w3c:wcag21}. Each principle is decomposed into success criteria (testable conditions) that span low-level syntactic checks (e.g., presence of \texttt{alt} attributes) to higher-level semantic and interactional behaviours (e.g., predictable keyboard focus order, timely and perceivable feedback for user actions, and reliable ARIA usage). Because WCAG is both normative and widely operationalized, it is the appropriate target for practical testing frameworks; however, its ambition—covering perception, interaction, comprehension and technical robustness—exposes a methodological challenge for automated tooling~\cite{standard:w3c:wcag21,standard:w3c:wcag22}. State-of-the-art automated tools - for example, axe-core, Lighthouse, and WAVE - perform an indispensable role by catching many low-level, syntactic violations. More recently, LLM-powered ones such as GenA11y~\cite{journal:gena11y:2025} showed improved capacity compared to traditional approach at catching accessibility error. Yet they largely operate under a static, page-centric evaluation model: tools evaluate a single rendered DOM snapshot or a finite set of page-load heuristics and then report rule-based findings or LLM-augmented decisions. 

This model is increasingly misaligned with modern web applications whose correctness and accessibility are inherently tied to \emph{sequences of interactions}, asynchronous updates, and transient UI states~\cite{conf:issta:ma11y:2024, conf:icse-seis:wave-history:2024, conf:w4a:accessibility-metatesting:2023}, i.e. \textbf{dynamic} accessibility issues. For instance, consider the following errors: a dynamically injected validation error that is not announced to assistive technologies and leaves focus on a non-descriptive element; a language toggle that updates visible text without updating the page \texttt{lang} attribute (so screen readers continue using the wrong language); or a dynamically created modal that neither manages focus nor exposes a programmatic label, producing a keyboard trap only at runtime. Together, such cases reveal a systematic coverage gap: snapshot-based audits often report syntactic compliance while overlooking interaction-dependent failures. None of the previously listed accessibility faults could be found using static information, since the evidence only occur at runtime. This motivates shifting from isolated page checks to \textbf{scenario-aware accessibility analysis}, which treats accessibility as an emergent property of multi-step user workflows.

To address this issue, we propose \toolname{}, a flow-aware agentic based system adapted for detecting a dynamic accessibility issue arising during a user interaction scenario. Rather than considering a static unit such as a loaded page, it works around dynamic scenario such as opening a menu, inputting a key etc. \toolname{} uses 
% \Florian{need to make clear how the scenarios are obtained}\Leisan{UPDATED}
user-defined scenarios and directly interacts with the web browser to perform said scenario.  Scenarios are manually curated natural-language task descriptions targeting real public websites, selected to cover interaction dependent WCAG 2.2 criteria that require live browser interaction to detect — such as keyboard operability (2.1.1, 2.1.2), focus management (2.4.3, 2.4.7), and status messages (4.1.3). During execution, runtime information are collected and are transformed into concrete evidence that can then be analyzed by the agent to support whether an accessibility issue occured or not during the scenario. Given that, to the best of our knowledge, there is no dataset of \textit{dynamic} WCAG, to support the evaluation of \toolname{} and future research,
% \Florian{need to make clear this part} \Leisan {UPDATED}
we also propose a dataset of 19 dynamic interaction scenarios covering 45 WCAG criterion checks across 19 real websites that were manually checked.
The anonymized artifact, including the source code, benchmark fixtures, oracle labels, prompt templates, executable notebook, and normalized result tables, is available for reproducibility~\cite{artifact:flow-a11y}.
We structure the evaluation around three research questions:

\begin{itemize}
    \item[\textbf{RQ1}] \textit{How accurately does \toolname{} assess WCAG-relevant accessibility outcomes that emerge during real user interaction flows?}
    \item[\textbf{RQ2}] \textit{How much does criterion-specific, runtime-evidence-grounded dynamic WCAG analysis improve accessibility assessment compared with a generic browser agent accessibility audit?}
    \item[\textbf{RQ3}] \textit{What classes of dynamic accessibility issues does \toolname{} identify reliably, and what failure modes currently limit its effectiveness?}
\end{itemize}

% \Florian{missing values: How much increase or raw accessibility issue found? compared to the baselin etc.}

Overall, the results show that criterion-specific runtime evidence is essential for assessing interaction-dependent WCAG behavior. Across 765 criterion assessments from successfully completed scenarios,  \toolname{} achieved 45.6\% exact five-way status accuracy, compared with 3.3\% for a generic browser-use accessibility audit over the same benchmark. The difference is particularly clear for failure detection: the baseline detected no oracle failures, whereas \toolname{} produced grounded fail findings with 41.4\% precision. 
% The final evidence-calibrated pipeline also eliminated invalid evidence references entirely, reducing them from 35 to 0, and reduced model calls from 609 to 203 relative to the pre-layer run.

Our analysis showed that criterion-specific runtime evidence and evidence-gated judgment are necessary for interaction-flow WCAG assessment, while exposing a precision-recall tradeoff whose implications for developer-facing accessibility tooling we analyse in the section \ref{sec:discussion}.

% \Florian{this can be removed if no space}
% We make the following contribution:
% \begin{itemize}
%     \item \toolname{}, an agent-based system for detecting dynamic WCAG accessibility barriers, encompassing manual-scenario execution, in-browser runtime evidence collection, criterion-specific probe analysis, LLM assissted judgment, and auditable artifact export.
%     \item A benchmark of 19 real interaction scenarios, 45 dynamic WCAG criteria, and 855 manually labelled scenario-criterion oracle rows, intended to support future research on interaction-flow accessibility assessment.
% \end{itemize}

%% file: sections/motivation.tex
% ============================================================
\section{Motivation and Problem Statement}
\label{sec:motivation}
% ============================================================

Web accessibility is not a property of a page in isolation.
Many barriers to access only become observable once a user begins to
interact: a search filter is applied, a modal dialog is opened, a menu
is expanded, a result list is updated by an asynchronous request, or a
keyboard user attempts to navigate newly loaded content.
These are not corner cases.
They are the normal conditions under which real users complete real tasks,
and they are precisely the conditions that snapshot-oriented accessibility
testing cannot evaluate.

\begin{figure}[t]
  \centering
  \begingroup
  \setlength{\fboxsep}{0pt}
  \setlength{\fboxrule}{0.35pt}

  \fbox{\includegraphics[width=0.49\columnwidth]
    {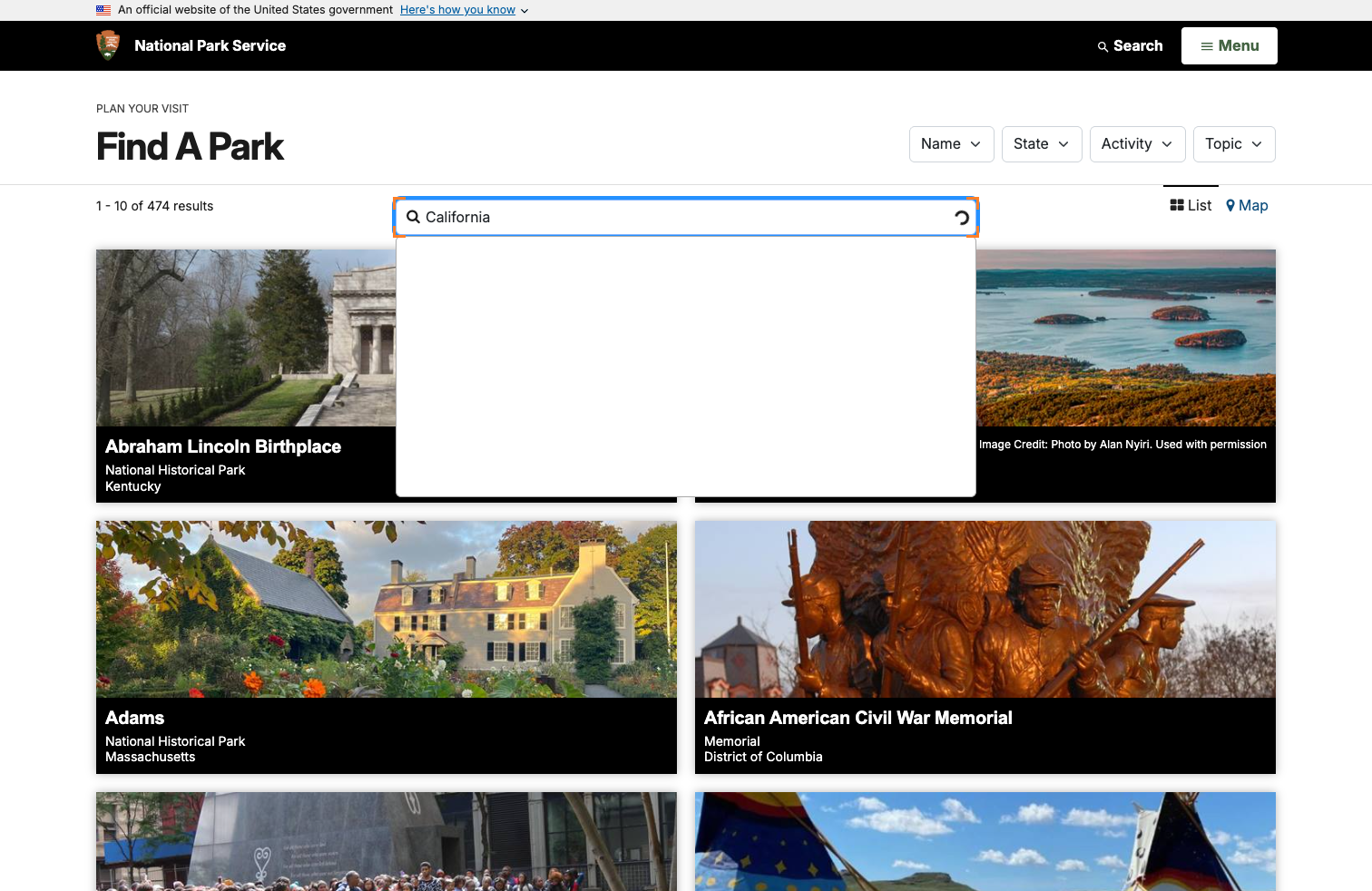}}
  \hfill
  \fbox{\includegraphics[width=0.49\columnwidth]
    {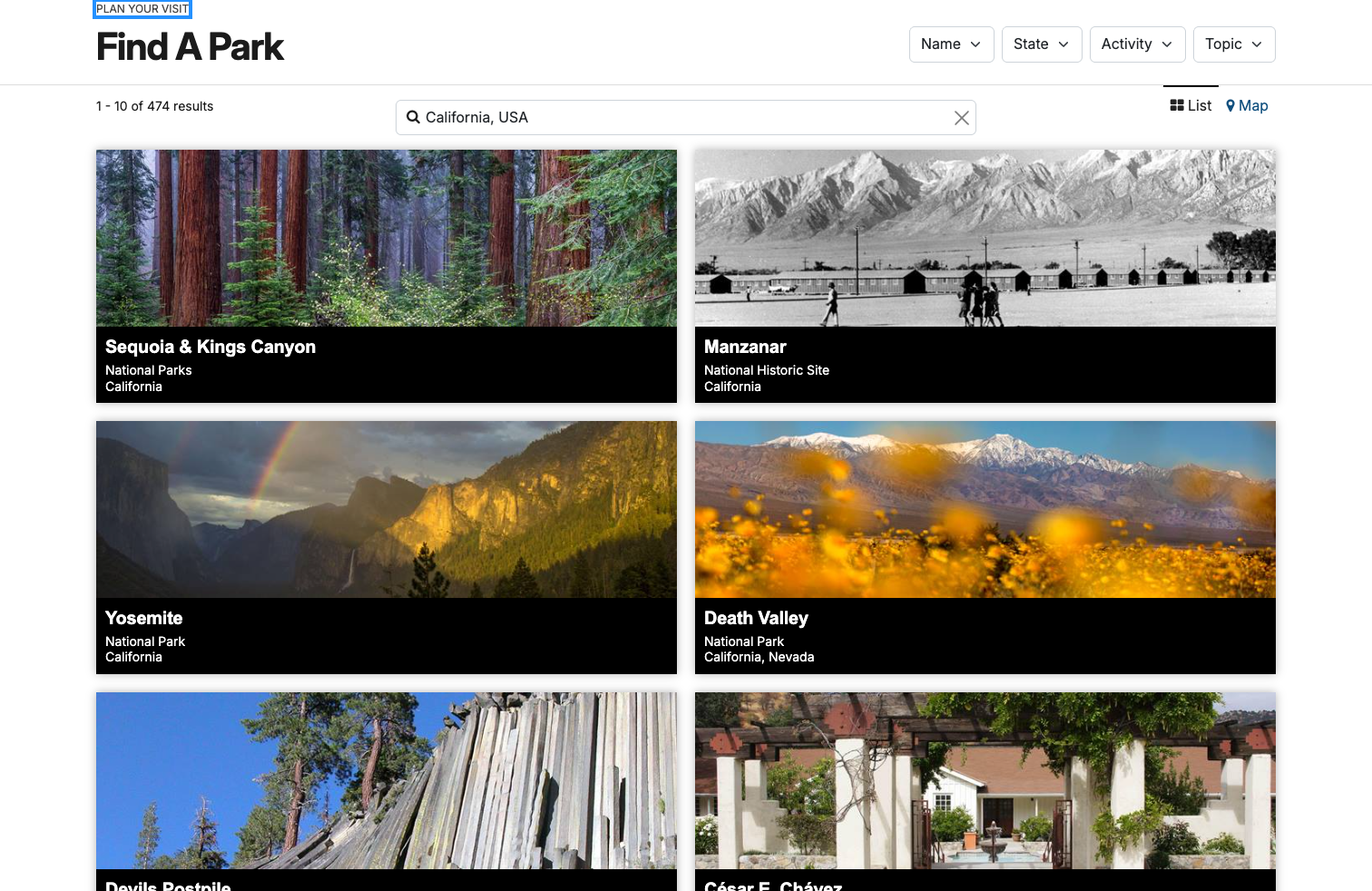}}

  \endgroup
  \caption{%
    A flow-aware accessibility barrier on the National Park Service
    \emph{Find A Park} page.
    \emph{Left:} the initial page state after loading.
    The search form is structurally present and visually intact.
    \emph{Right:} the state reached after a user searches for parks in
    California and begins keyboard-navigating the updated result list.
    Whether keyboard focus remains visible and traverses the result cards
    in a logical order is a WCAG question that depends entirely on this
    runtime state -- it cannot be answered from the initial snapshot.
  }
  \label{fig:motivating-flow}
\end{figure}

\Cref{fig:motivating-flow} illustrates the problem with a concrete
scenario drawn from our benchmark. A user visits the National Park Service \emph{Find A Park} page (www.nps.gov), enters a location, and attempts to navigate the resulting park list using the keyboard.
At the moment the page loads (\Cref{fig:motivating-flow}, left), a
conventional automated checker can confirm that the search form has a
label, that input fields are reachable, and that the visible text has
sufficient contrast.
These observations are correct and useful.
What they cannot answer is whether, after the search executes and the
result list updates (\Cref{fig:motivating-flow}, right), keyboard focus
moves into the new content in a predictable order, remains visually
discernible against the result card backgrounds, and does not become
trapped.
Those properties depend on the sequence of interactions that produced the
current browser state.

This is not a limitation unique to any particular tool.
It reflects the unit of analysis that rule-based scanners use: the
observable DOM at a single point in time.
WCAG 2.2 includes a family of success criteria whose applicability is
inherently interaction-dependent.
Criteria such as \emph{Keyboard} (SC~2.1.1), \emph{No Keyboard Trap}
(SC~2.1.2), \emph{Focus Order} (SC~2.4.3), \emph{Focus Visible}
(SC~2.4.7), \emph{Focus Appearance} (SC~2.4.13), \emph{Content on Hover
or Focus} (SC~1.4.13), \emph{Status Messages} (SC~4.1.3),
\emph{Timing Adjustable} (SC~2.2.1), and \emph{Dragging Movements}
(SC~2.5.7) all require observations that span user actions, focus
transitions, state updates, and DOM mutations~\cite{standard:w3c:wcag22}.
A judgment on any of these criteria cannot be derived from a static page
audit; it requires evidence gathered during an interaction.

We define this problem as \emph{flow-aware accessibility testing}: the
task of assessing whether WCAG-relevant barriers arise at the runtime
states reached during a user interaction scenario.
A flow-aware tester must execute the scenario in a real browser, capture
runtime evidence, which includes focus transitions, keyboard events, DOM
mutations, accessibility-tree snapshots, screenshots, and timing behavior, and then evaluate WCAG criteria whose applicability depends on that accumulated evidence.

%% file: sections/background.tex
\section{Background and Related Work}
\label{sec:background}

% ============================================================

\subsection{WCAG and accessibility conformance}
The Web Content Accessibility Guidelines (WCAG) define technology-neutral
success criteria organized around four principles: content must be
perceivable, operable, understandable, and robust~\cite{standard:w3c:wcag22}.
WCAG 2.2 extended the standard with criteria particularly relevant to
interaction, including \emph{Focus Not Obscured} (SC~2.4.11--12),
\emph{Focus Appearance} (SC~2.4.13), \emph{Dragging Movements}
(SC~2.5.7), \emph{Consistent Help} (SC~3.2.6), \emph{Redundant Entry}
(SC~3.3.7), and \emph{Accessible Authentication} (SC~3.3.8--9).
Conformance evaluation typically combines automated scanning,
semi-automated review, and manual testing with assistive technology,
as reflected in W3C evaluation guidance and WCAG-EM~\cite{standard:w3c:wcag21,standard:w3c:wcag22,standard:w3c:wcag-em}.
Automated tools are a necessary component of this process, but their
coverage is bounded by what is statically observable.

Independent analyses estimate that rule-based automated scanners reliably
detect approximately 30--57\% of real-world WCAG
failures~\cite{web:deque:axe-core,conf:fse:accessitext:2022,report:deque:automated-coverage,conf:w4a:accessibility-metatesting:2023}.

\subsection{Rule-based accessibility checkers}
Tools such as axe-core~\cite{web:deque:axe-core}, WAVE~\cite{web:webaim:wave},
Lighthouse~\cite{web:google:lighthouse}, and Pa11y~\cite{web:pa11y:pa11y}
are the primary means by which accessibility is validated in development
workflows today.
These tools evaluate accessible names, ARIA roles, color contrast,
landmark structure, and form labeling by inspecting the DOM at the time
of the audit.
They are valuable, scalable, and widely
integrated~\cite{web:deque:axe-core,web:google:lighthouse}.
However, as their own documentation acknowledges, they are organized around page-level audits over observable page state.
A static DOM scan can confirm that a modal dialog carries a role="dialog" attribute
and an accessible name; it cannot verify that keyboard focus moves into the dialog
when it opens, that focus is contained within it during navigation, or that focus
returns correctly when it closes~\cite{standard:w3c:aria-apg-dialog,conf:chi:bagel:2023}.
These behaviors depend on user interaction and are outside the scope of snapshot-based analysis.
\toolname{} is orthogonal to these tools rather than a replacement for them: it targets the class of WCAG issues that snapshot-based analysis cannot reach by construction.

\subsection{LLM-based accessibility analysis}
Recent work has demonstrated that large language models can extend automated
accessibility analysis beyond syntactic rules.
GenA11y~\cite{journal:gena11y:2025} extracts criterion-relevant page elements and prompts an LLM to detect WCAG violations, reporting detection of additional violations per page that rule-based tools miss~\cite{journal:gena11y:2025}.
L\'opez-Gil and Pereira~\cite{journal:llm-manual-wcag-criteria:2025} show that
LLM-based scripts can automate WCAG criteria that traditionally require manual evaluation.
Other work investigates LLM-assisted remediation~\cite{arxiv:accessguru:2025}
and code generation for accessibility
compliance~\cite{arxiv:accessible-code-generation:2025}.
These systems show that semantic reasoning over page content can improve
WCAG coverage.
\toolname{} builds on this direction but differs in its primary input:
rather than a loaded page or its DOM, \toolname{} operates on a runtime
trace accumulated during a user interaction scenario.
Criterion-specific probes determine what evidence must be present before an
LLM analysis is triggered, and the analyzer is gated on that evidence rather
than on page structure alone.

\subsection{Interaction-aware and dynamic accessibility testing}
Researchers have begun to address the limitations of snapshot-based
accessibility testing in specific interaction contexts.
Chiou~et~al.~\cite{conf:fse:keyboard-accessibility-failures:2021} detect keyboard
accessibility failures in web applications by driving focus-related
interactions with a browser agent.
Their subsequent work targets dialog-related
failures~\cite{conf:icse:dialog-keyboard-navigation:2023} and reflow
issues~\cite{conf:icse:reflow-accessibility:2024}, each narrowly defined for a
specific class of barrier.
On the mobile side, Mehralian~et~al.~\cite{conf:icse:timestump:2025}
analyze dynamic content changes in mobile apps, detecting
accessibility issues that emerge during rendering transitions rather than
at stable screen states.
Salehnamadi~et~al.~\cite{conf:chi:a11ypuppetry:2023} use record-and-replay
to support assistive-technology-aided manual testing of interaction flows in
mobile apps. However, these approaches are limited to specific scenario or targeted WCAG dynamic issues, where prior work addresses specific barrier classes in isolation, \toolname{} provides a unified framework for criterion-specific, evidence-gated analysis across a set of interaction-dependent
WCAG criteria, currently covering 45 dynamic success criteria spanning 9 probe families.

% \Florian{This does not fit here}
% \toolname{} is not intended to replace snapshot-based accessibility checkers.
% Instead, it targets a complementary class of barriers: WCAG-relevant failures
% whose evidence appears only after executing an interaction flow.
% Where prior web-focused work addresses specific barrier classes (keyboard
% traps, dialog focus, reflow), \toolname{} provides a unified framework for
% criterion-specific, evidence-gated analysis across a broader set of
% interaction-dependent WCAG criteria.
% It does so by pairing each criterion with the runtime evidence required to
% assess it -- focused element state, keyboard events, DOM mutations,
% accessibility-tree snapshots, screenshots, timing records, and media state
% -- and by executing user-defined interaction scenarios to collect that
% evidence in a real browser.
% Existing tools are valuable for scalable page-level checks; \toolname{}
% addresses the runtime trace that those tools, by design, do not capture.

%% file: sections/approach.tex
\section{Approach}
\label{sec:approach}

% ============================================================
\subsection{\toolname{} Overview}
\label{subsec:overview}
% ============================================================

\toolname{} is a flow-aware accessibility testing system that evaluates
WCAG-relevant barriers emerging during user interaction scenarios.
Its central design decision is to treat an \emph{interaction scenario},
rather than a loaded page, as the unit of accessibility assessment.
A scenario specifies a concrete user task -- opening a disclosure menu,
submitting a form with invalid input, keyboard-navigating a search result
list, expanding an accordion, or closing a modal dialog -- and
\toolname{} executes that task in a real browser, records the runtime
states reached during the interaction, and analyzes those states against
dynamic WCAG criteria.

\begin{figure*}[t]
  \centering
  \includegraphics[
    width=0.60\textwidth
  ]{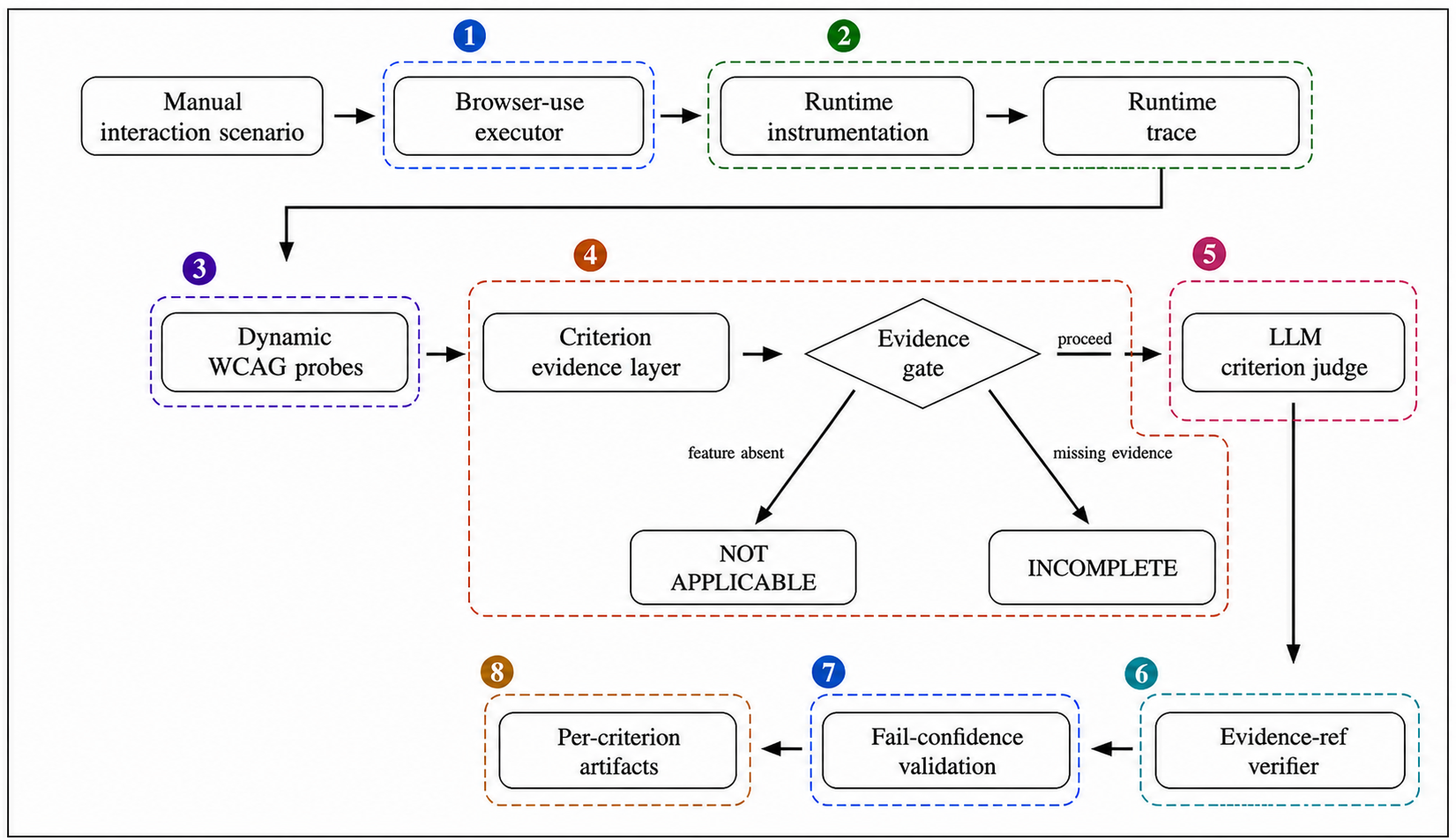}
  \caption{\toolname{} end-to-end architecture. The system executes a scenario, captures a runtime trace, projects it into criterion-specific evidence, gates unsupported judgments before model invocation, verifies model evidence references, and exports auditable criterion-level artifacts.}
  \label{fig:a11yextreme-architecture}
\end{figure*}

\Cref{fig:a11yextreme-architecture} illustrates the full analysis pipeline.
In the stage~\stageone, a \emph{browser-use executor} performs the scenario
task and captures the browser state before and after each user action.
In the stage~\stagetwo, \emph{runtime extractors} collect multimodal
evidence from each state: the DOM, accessibility tree, currently focused
element, keyboard events, screenshots, DOM mutations, timing signals,
media state, and console output.
Together, the ordered sequence of states and their evidence form a
\emph{runtime trace} for the scenario.
In the stage~\stagethree, \emph{dynamic probes} transform the raw trace into
criterion-oriented evidence structures: focus sequences, focus-visibility
observations, status-message records, hover and focus content snapshots,
pointer-interaction indicators, and form-error feedback.
In the stage~\stagefour, a \emph{criterion evidence layer} constructs a
compact, family-specific evidence packet for each WCAG criterion,
assesses applicability, evaluates evidence quality, and applies a
pre-LLM gate: criteria with absent or inapplicable evidence are
resolved without invoking the model.
In the stage~\stagefive, criteria that pass the gate proceed to
\emph{LLM-based criterion judgment} using criterion-specific prompts
that include the evidence packet, the WCAG criterion definition, and a
structured output schema.
In the stage~\stagesix, an \emph{evidence-reference verifier} checks that
\textsc{fail} and \textsc{warning} outputs cite resolvable fields from
the evidence packet; model outputs that cite non-existent fields are
downgraded to \textsc{incomplete}.
In the stage~\stageseven, a \emph{fail-confidence validator} inspects
whether cited evidence is direct and behavioral or only indirect and
heuristic; fail outputs grounded solely in indirect evidence are
downgraded to \textsc{warning}.
Finally~\stageeight, a \emph{calibrated status router} assigns each result a
decision level -- \emph{confirmed\_fail}, \emph{probable\_violation},
\emph{weak\_signal}, \emph{insufficient\_evidence}, or
\emph{not\_applicable} -- and the system exports a structured artifact package for each criterion.

\subsection{Scenario-Aware Dynamic Accessibility Assessment}
\label{subsec:scenario}
% ============================================================

% \Florian{How are scenario obtained? It needs to be said}
The scenario is the fundamental test artifact in \toolname{}.
A scenario specifies a user goal and the web application context required
to pursue it. It is not limited to an initial URL: a scenario may require the agent to search, apply a filter, open a menu, trigger a form-validation error, navigate loaded content with the keyboard, interact with media controls, or wait for asynchronous feedback. The purpose of the scenario is to drive the browser into the states in which interaction-dependent accessibility barriers can occur.

\paragraph{Scenario execution and trace formation}

For each scenario, \toolname{} launches a browser-use executor that performs the task and records an ordered sequence of runtime states.
The sequence begins with an initial state capturing the browser context before any agent action, and continues with a post-action state after each browser-use step.
The complete trace represents not what the page looked like at load time, but the trajectory of browser state that the user interaction produced. The precise evidence fields captured at each state are described in \Cref{subsec:evidence}.

\paragraph{Why scenario execution is necessary}
The interaction-dependent WCAG criteria introduced in \Cref{sec:motivation} share a common property: their applicability cannot be determined from a static page audit. Many WCAG criteria are only assessable after an interaction has occurred.

Keyboard operability (SC~2.1.1) requires observing how focus moves during keyboard use, not only whether interactive elements are markup-reachable.
Focus visibility (SC~2.4.7, SC~2.4.11) requires evidence about the element that actually received focus during a keyboard sequence, not only whether a focus style exists in the stylesheet. Status messages (SC~4.1.3) require triggering the event that produces the message, then observing how assistive-technology-relevant markup reflects
it.
Form-error criteria (SC~3.3.1, SC~3.3.3) require observing validation feedback after user input is submitted.
Timing, pointer, hover/focus-content, dragging, and authentication criteria similarly depend on runtime behavior whose presence or absence cannot be
determined from a static page audit.
A scenario that executes the relevant interaction, and a trace that records the resulting states, are the minimal evidence base for assessing any of these criteria.

\paragraph{Separation of execution and assessment}
\toolname{} separates task execution from criterion assessment.
The browser executor is responsible for reaching the runtime states that expose interaction-dependent behavior.
The dynamic probes and WCAG analyzer, described in
\Cref{subsec:probes,subsec:overview}, are responsible for interpreting those states against specific WCAG criteria.
This separation has two practical consequences: First, the same runtime trace supports multiple criterion judgments
simultaneously; re-running the scenario for each criterion is not required. Second, the evaluation distinguishes genuine accessibility judgments from \textsc{incomplete} results caused by the absence of required evidence -- for example, a keyboard-focus criterion that cannot be assessed because no keyboard event occurred during execution.
This distinction is essential for interpreting evaluation results honestly, and for guiding future scenario design toward the states that specific criteria require.

% ============================================================

% ============================================================

% ============================================================

% ============================================================
\subsection{Runtime Evidence Model}
\label{subsec:evidence}
% ============================================================

% \Florian{This seems more like something in the Methodology. First, you should have here your benchmark} --- UPDATED
\toolname{} represents each accessibility assessment as a
\emph{runtime trace} rather than as an isolated page snapshot. This design choice reflects the nature of the WCAG criteria being evaluated: interaction-dependent barriers do not exist at page-load time; they arise at the runtime states produced by user actions.
Preserving an ordered record of those states is therefore a prerequisite for any criterion judgment that requires observing focus movement, keyboard operation, DOM mutation, live-region updates, or media behavior.

\paragraph{Trace structure}
A runtime trace consists of scenario metadata and an ordered sequence of \emph{runtime states}.
The first state is captured before any agent action and records the initial browser context.
Subsequent states are captured after each browser-use step.
Each state records: the page URL and title, the action context that produced it, a DOM snapshot, an accessibility-tree snapshot, the currently focused element and its accessible properties, keyboard event records,
console output, a screenshot reference, DOM mutation logs, timing signals, media playback state, and interaction metadata. The ordered sequence of states encodes the trajectory of browser behavior that the scenario produced, rather than any single moment within it.

\paragraph{Browser instrumentation}
Before execution begins, \toolname{} installs JavaScript event collectors in the browser.
These collectors listen for keyboard, pointer, focus, input, form submission, live-region mutation, and media events. At each step, the collector streams are drained and written as step-level artifacts alongside the structural and visual evidence. The resulting trace therefore combines four evidence types simultaneously:
\emph{structural evidence} (DOM, accessibility tree),
\emph{visual evidence} (screenshots), \emph{behavioral evidence} (keyboard events, focus transitions, DOM mutations, pointer activity), and \emph{state-change evidence} (timing records, media state, live-region updates, console signals). No single evidence type is sufficient in isolation for the criteria \toolname{} addresses; the combination is what enables criterion-specific gating in the analyzer.

\paragraph{Design rationale}
The trace model is designed around two requirements. First, evidence must be \emph{temporally ordered}: a focus-visibility judgment depends on which element received focus and when, not only on whether a CSS focus style exists. A status-message judgment depends on the DOM  mutation that created the message and on the accessibility-tree annotation at that moment. Temporal ordering preserves these dependencies. Second, evidence must be \emph{criterion-addressable}: the analyzer must be able to extract the specific fields relevant to a given WCAG criterion without reprocessing the full trace. The schema is therefore organized to support per-criterion evidence projection, which the probe layer uses to construct criterion-specific input bundles for the WCAG analyzer.

% ============================================================
\subsection{Dynamic WCAG Probes and Analyzer}
\label{subsec:probes}
% ============================================================

The probe and analyzer layer translates a raw runtime trace into
criterion-level accessibility judgments through a criterion-calibrated
decision pipeline.
Rather than passing raw probe evidence directly to a language model,
the pipeline first constructs a family-specific evidence packet for
each criterion, gates model invocation on evidence sufficiency and
applicability, verifies that model outputs cite resolvable runtime
evidence, and applies a post-judgment confidence check before routing
each result to a calibrated decision level.
This design separates what the model is allowed to reason about from
what it is allowed to assert.

\subsubsection{WCAG Registry and Probe Mapping}
\label{subsec:probes:registry}
A WCAG registry maps each of the 45 supported dynamic criteria to the
probe class responsible for extracting its evidence and to the specific
evidence fields required for assessment.
The registry drives both probe execution and evidence gating: a criterion
cannot proceed to judgment unless the fields its registry entry requires
are present in the trace.
Nine runtime probes cover the full set of supported criteria,
summarised in Table~\ref{tab:dynamic-wcag-families}, and detailed WCAG criteria are all listed in our replication package.
Probes for keyboard, focus, live-region, media, and event-listener
behavior are grounded directly in the instrumentation streams described
in Section~\ref{subsec:evidence}.
Probes for timing, animation, and consistency rely on a combination of
runtime signals and DOM heuristics; we note this distinction where
relevant in the evaluation.

\begin{table}[H]
\centering
\small
\caption{Dynamic WCAG criterion families in \toolname{}.}
\label{tab:dynamic-wcag-families}
\setlength{\tabcolsep}{4pt}
\begin{tabular}{p{0.30\columnwidth}p{0.62\columnwidth}}
\toprule
\textbf{Probe family} & \textbf{Runtime behavior captured} \\
\midrule
Keyboard \& focus        & Focus sequences, keyboard events, tab order, focus containment \\
Hover \& focus content   & Content appearing/disappearing on hover or keyboard focus \\
Timing \& interruption   & Time limits, interruption signals, re-authentication timeouts \\
Animation \& flashing    & Flashing events, motion, reduced-motion behavior \\
Pointer \& gesture       & Pointer handlers, drag paths, cancellation, motion actuation \\
Consistency \& navigation & Navigation/identification snapshots, help, URL sequences \\
Forms, errors \& auth    & Input/submit events, error messages, authentication signals \\
Status messages          & Live-region mutations, role/status/alert signals \\
Media                    & Playback state, caption tracks, audio-description indicators \\
\bottomrule
\end{tabular}
\end{table}

\subsection{Criterion Evidence Packets}
\label{subsec:probes:packet}

Before invoking the judgment model, the analyzer constructs a
\emph{criterion evidence packet} for each criterion that passes the
applicability gate.
The packet is family-specific: the builder applies distinct logic for
each probe class.
For \emph{keyboard and focus} criteria, the packet requires keyboard
event records, focus sequences, before/after focus element transitions,
and visible focus candidate observations.
For \emph{media} criteria, the packet is populated only when media is
confirmed to be playing, time-advancing, or audibly active in the trace.
For \emph{pointer} criteria, passive event-listener presence alone is
insufficient; the packet requires task-relevant drag path, motion, or
multi-pointer evidence.
For \emph{form} criteria, generic error text is treated as weak evidence;
the packet requires input, submit, or change events tied to visible or
programmatic feedback.
Each packet records the criterion family, a binary applicability flag
with reason, evidence strength, lists of positive, missing, and
contradictory evidence items, step consistency across trace states, and
compact runtime facts projected from the full trace.
This packet replaces the raw evidence bundle as the primary input to
the judgment model, constraining the model to family-relevant signals.

\begin{figure}[t]
\centering
\setlength{\fboxsep}{3pt}
\fbox{%
\begin{adjustbox}{max width=0.88\columnwidth}
\begin{tikzpicture}[
  font=\tiny,
  node distance=0.28cm,
  block/.style={draw, rounded corners, align=center, minimum height=0.42cm, minimum width=2.15cm, inner sep=1.5pt},
  smallblock/.style={draw, rounded corners, align=center, minimum height=0.42cm, minimum width=1.85cm, inner sep=1.5pt},
  gate/.style={draw, diamond, aspect=2.0, align=center, inner sep=0.5pt, minimum width=1.15cm},
  arr/.style={-{Latex[length=1.4mm]}, thick},
  lab/.style={font=\tiny, fill=white, inner sep=0.5pt}
]

\node[block] (probe) {Probe evidence\\for one criterion};
\node[block, below=0.18cm of probe] (req) {Required-evidence\\validator};
\node[block, below=0.18cm of req] (app) {Applicability\\analysis};
\node[block, below=0.18cm of app] (quality) {Evidence-quality\\summary};
\node[block, below=0.18cm of quality] (packet) {Criterion evidence\\packet};
\node[gate, below=0.18cm of packet] (gate) {Gate\\decision};

\node[smallblock, below left=0.32cm and 0.85cm of gate] (na) {Return\\NOT APPLICABLE};
\node[smallblock, below=0.52cm of gate] (inc) {Return\\INCOMPLETE};
\node[smallblock, below right=0.32cm and 0.85cm of gate] (prompt) {Render criterion\\prompt};

\node[block, below=0.38cm of prompt] (gemini) {LLM};
\node[block, below=of gemini] (verify) {Resolve cited\\evidence refs};
\node[block, below=of verify] (downgrade) {Accept or\\downgrade output};

\draw[arr] (probe) -- (req);
\draw[arr] (req) -- (app);
\draw[arr] (app) -- (quality);
\draw[arr] (quality) -- (packet);
\draw[arr] (packet) -- (gate);

\draw[arr] (gate) -- node[lab, above left, pos=0.48] {absent} (na);
\draw[arr] (gate) -- node[lab, right, pos=0.48] {missing} (inc);
\draw[arr] (gate) -- node[lab, above right, pos=0.48] {proceed} (prompt);

\draw[arr] (prompt) -- (gemini);
\draw[arr] (gemini) -- (verify);
\draw[arr] (verify) -- (downgrade);

\end{tikzpicture}
\end{adjustbox}%
}
\caption{Pre-LLM evidence layer: unsupported criteria are resolved before model invocation, and model findings must cite resolvable evidence.}
\label{fig:pre-gemini-evidence-layer}
\end{figure}
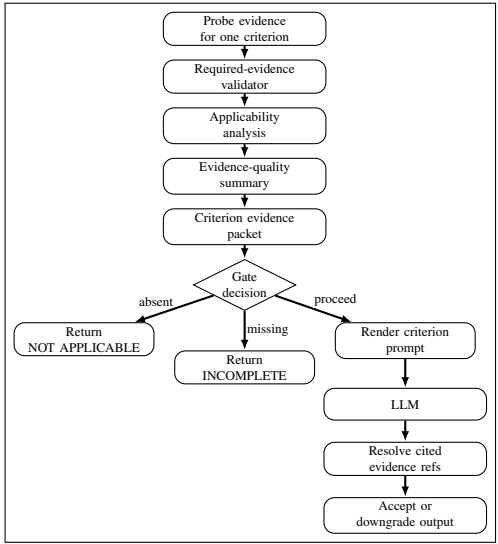

% ============================================================

\subsection{Evidence-Gated LLM Judgment}
\label{subsec:llm-judgment}

\subsubsection{Evidence gating}
Before any criterion judgment is produced, the analyzer applies an evidence
gate.
If the required evidence fields for a criterion are absent from the trace
-- because the scenario did not trigger the relevant interaction -- the
result is marked \textsc{incomplete}.
If the evidence is present but indicates that the criterion is not
applicable to the current scenario, for example because no media element
exists on any visited page, the result is marked \textsc{not\_applicable}.
Only criteria with sufficient and applicable evidence proceed to judgment.
This discipline prevents the analyzer from producing spurious judgments
in the absence of the interactions that a criterion requires.

\subsubsection{Criterion judgment}
In the publication evaluation, all criteria that pass the evidence gate
are judged using an LLM with criterion-specific prompts.
The prompt construction and output contract are described below.

\subsubsection{Prompt construction and grounded output contract}
\toolname{} uses criterion-specific prompts generated from static templates
and runtime evidence.
Each prompt states the WCAG criterion, the responsible probe, required
evidence keys, allowed statuses, and a strict JSON schema.
The analyzer then injects the criterion evidence payload and soft
evidence-quality guidance before invoking the LLM.
The prompt contract requires \textsc{fail} and \textsc{warning} outputs
to cite concrete \texttt{evidence\_refs}; after model inference, the
analyzer mechanically resolves these references against the evidence JSON
and downgrades unsupported outputs to \textsc{incomplete}.
This design makes the model a criterion judge over bounded runtime
evidence rather than a free-form accessibility reviewer.

\begin{figure}[t]
\centering

\setlength{\fboxsep}{6pt}

\begin{adjustbox}{max width=\columnwidth}
\begin{minipage}{\columnwidth}
\scriptsize
\begin{rqresultbox}
\begin{verbatim}

Criterion: 2.4.3 Focus Order
Probe: keyboard_focus_probe
Required evidence: focus_sequence
Rules:
  - Evidence JSON is the only source of truth.
  - Do not invent user actions, DOM nodes, screenshots,
  or focus behavior.
  - Fail only with direct evidence of non-compliance.
  - Use warning for plausible but indirect risk.
  - Each fail/warning violation must cite resolvable
  evidence_refs.
Output:
  {criterion_id, status, summary, missing_evidence,
  violations[]}

\end{verbatim}
\end{rqresultbox}

\end{minipage}

\end{adjustbox}

\caption{Abbreviated prompt contract for a focus-order judgment. The full prompt is generated from a criterion-specific template and the runtime evidence packet for the scenario.}
\label{fig:prompt-contract}
\end{figure}
\subsubsection{Fail-Confidence Validation}
\label{subsec:probes:fail-confidence}

After the LLM judgment, \textsc{fail} results undergo a deterministic
post-processing step.
The \emph{fail-confidence validator} inspects whether the evidence cited
by the model is direct and behavioral -- for example, a keyboard event
confirming focus loss, or a DOM mutation confirming a missing live-region
announcement -- or indirect and heuristic -- for example, the absence of
an ARIA attribute in the initial DOM without corroborating behavioral
evidence.
\textsc{fail} outputs grounded only in indirect or absent evidence are
downgraded to \textsc{warning}.
This step does not modify \textsc{pass}, \textsc{incomplete}, or
\textsc{not\_applicable} outputs.
This filter reduced the number of invalid evidence references in the result tables from 35 to~0, and improved
fail precision relative to a run without the evidence layer. The cases that move from \textsc{fail} to \textsc{warning} are
analysed in \Cref{sec:results:confusion}.

\subsubsection{Calibrated Status Routing}
\label{subsec:probes:routing}

The final output of the analyzer is not a binary pass/fail label but a
\emph{decision level} drawn from five categories:
\emph{confirmed\_fail} (direct behavioral evidence of a WCAG violation),
\emph{probable\_violation} (strong but not fully direct evidence),
\emph{weak\_signal} (indirect evidence insufficient for a hard failure
claim),
\emph{insufficient\_evidence} (evidence gate not met), and
\emph{not\_applicable} (criterion irrelevant to this scenario).
For primary metric reporting, \textsc{fail} corresponds to
\emph{confirmed\_fail} only.
\textsc{warning} encompasses \emph{probable\_violation} and
\emph{weak\_signal}.
This distinction matters for developers consuming the output: a
\emph{confirmed\_fail} is a grounded, artifact-backed finding requiring
remediation, while a \textsc{warning} is a signal requiring human review.
The precision-recall tradeoff between these tiers is a structural
consequence of evidence calibration and is analysed in
\Cref{sec:results}.

% ============================================================

\subsection{Auditable Outputs}
For each criterion evaluated, \toolname{} exports a structured artifact
package containing: the criterion-level result and decision level, the
criterion evidence packet, the evidence gate decision and its reason,
the fail-validation outcome where applicable, the prompt sent to the
model, the model response, and -- where a screenshot was cited -- a
copy of the referenced image.
This makes every criterion decision auditable: a developer inspecting a
reported failure can review the exact evidence the analyzer used, the
family-specific packet that grounded the model invocation, the gate
decision that preceded it, and the interaction state from which the
evidence was drawn.
Auditability is particularly important for dynamic accessibility
testing, where relevant evidence may have appeared transiently during
a scenario and may not be observable from the page in its current state.
We found that all 201 model calls produced valid evidence-reference
outputs; zero invalid reference artifacts were recorded in the final
result tables.

% ============================================================

% ============================================================

%% file: sections/experimental_design.tex
\section{Experimental Design}
\label{sec:design}

% ============================================================
\subsection{Benchmark Design}
\label{subsec:benchmark}
% ============================================================

% \Florian{motivation?}

Existing accessibility benchmarks and evaluations primarily target page-level
snapshots, synthetic mutations, or individual barrier classes
~\cite{journal:gena11y:2025,conf:issta:ma11y:2024,conf:chi:bagel:2023}.
To the best of our knowledge, there is no benchmark combining executable multi-step interactions on
public websites, criterion-level oracle labels, and runtime evidence across a
broad set of interaction-dependent WCAG criteria. Because such a benchmark is
necessary to evaluate accessibility at the level of user flows, we constructed
one specifically for this study.

The benchmark evaluates \toolname{} on real interaction scenarios executed
against public websites.
Its design reflects three requirements: scenarios must exercise the
interaction types that dynamic WCAG criteria require; oracle labels must
cover the full criterion-by-scenario matrix at publication quality; and
the evaluation must be executable from a well-defined run configuration
with no oracle leakage into generated artifacts.

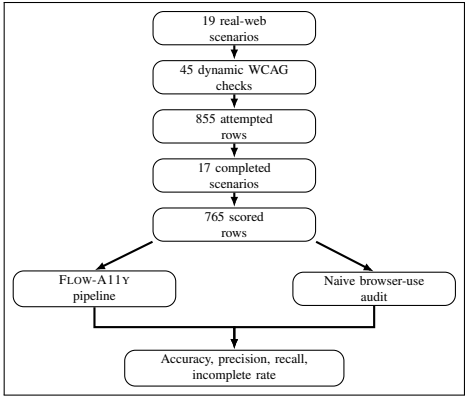
\begin{figure}[H]
\centering
\setlength{\fboxsep}{3pt}
\fbox{%
\begin{adjustbox}{max width=0.88\columnwidth}
\begin{tikzpicture}[
  font=\tiny,
  block/.style={draw, rounded corners, align=center, minimum height=0.44cm, minimum width=2.15cm, inner sep=1.5pt},
  wideblock/.style={draw, rounded corners, align=center, minimum height=0.44cm, minimum width=2.9cm, inner sep=1.5pt},
  arr/.style={-{Latex[length=1.4mm]}, thick}
]

\node[block] (scenarios) at (0, 3.70) {19 real-web\\scenarios};
\node[block] (criteria)  at (0, 3.05) {45 dynamic WCAG\\checks};
\node[block] (attempted) at (0, 2.40) {855 attempted\\rows};
\node[block] (completed) at (0, 1.75) {17 completed\\scenarios};
\node[block] (scored)    at (0, 1.10) {765 scored\\rows};

\node[block]     (a11y)    at (-1.85, 0.25) {\toolname{}\\pipeline};
\node[block]     (base)    at (1.85, 0.25) {Naive browser-use\\audit};
\node[wideblock] (metrics) at (0, -0.8) {Accuracy, precision, recall,\\incomplete rate};

\draw[arr] (scenarios) -- (criteria);
\draw[arr] (criteria) -- (attempted);
\draw[arr] (attempted) -- (completed);
\draw[arr] (completed) -- (scored);

\draw[arr] (scored.south west) -- (a11y.north);
\draw[arr] (scored.south east) -- (base.north);

\draw[arr] (a11y.south) -- ++(0,-0.263) -- ++(1.85,0) -- (metrics.north);
\draw[arr] (base.south) -- ++(0,-0.28) -- ++(-1.85,0) -- (metrics.north);

\end{tikzpicture}
\end{adjustbox}%
}
\caption{Benchmark design for the primary comparison.}
\label{fig:benchmark-design}
\end{figure}

\paragraph{Scenarios and website selection}

We constructed the benchmark by first enumerating interaction families needed by the 45 supported dynamic WCAG checks, we then selected publicly reachable websites containing representative instances of these interaction patterns. We excluded sites requiring accounts, CAPTCHAs, special permissions, or flows that could not be consistently reached by a browser agent. For each retained website, we wrote a natural-language task specifying the target URL, user goal, and interaction sequence needed to expose the relevant runtime state. The resulting benchmark contains 19 manually curated public-web scenarios. It is designed for dynamic WCAG coverage and external realism, not as a
random sample of websites.

\paragraph{Oracle construction}
The benchmark defines a $19 \times 45$ oracle matrix covering all
scenario-criterion combinations, yielding 855 oracle rows.
Each oracle entry records the expected criterion outcome for the given
scenario: \textsc{pass}, \textsc{fail}, \textsc{not\_applicable}, or
\textsc{warning}.
Oracle labels were established through \textit{manual review} of the target
applications by the authors.
Each entry includes a reviewer field, a confidence level, and optional
notes documenting the rationale for ambiguous cases.
A leakage check is performed at benchmark runtime: the runner scans
generated trace artifacts and reports for oracle field values before
computing publication metrics.
No oracle values are injected into prompts or trace artifacts at any point during execution.

\paragraph{Execution and scoring}
Each scenario is executed once using a browser-use agent.
\toolname{} captures the runtime trace, runs the dynamic WCAG analyzer,
and exports per-criterion results and artifact packages.
Of the 19 scenarios attempted, 17 completed successfully; 2 scenarios
encountered browser-task execution failures and produced no analyzable
trace.
The primary evaluation therefore covers 765 scored rows ($17 \times 45$).
The 90 rows from failed scenarios are excluded from all reported primary
metrics and discussed separately in \ref{sec:discussion}.
All reported accuracy, precision, recall, and incomplete-rate figures are
computed over the 765 scored rows only.

\paragraph{Baseline}
As, to the best of our knowledge, there is no existing system addressing the same flow-level,
criterion-specific assessment task, we constructed a naive browser-use global
audit baseline. The agent receives the target URL, a step budget, a general
instruction to explore the website for accessibility concerns, and the
supported WCAG criterion identifiers. It returns structured criterion-level
statuses, with unassessed criteria normalized to \textsc{incomplete}. Unlike
\toolname{}, it receives no interaction scenario, criterion-specific probes,
evidence packets, or evidence requirements. The baseline therefore serves as
both a naive browser-agent comparison and a coarse system-level ablation of
\toolname{}'s structured analysis pipeline, rather than a controlled ablation
of a single component.

\paragraph{Tool choices}
The pipeline components were selected to support scenario-level accessibility assessment. Since the evaluation unit is an interaction flow, \toolname{} uses \texttt{browser-use} to execute natural-language tasks and expose action-level browser states without hand-written per-site scripts~\cite{web:github:browser-use:2024}. Because the targeted WCAG criteria depend on evidence distributed across the interaction---focus transitions, keyboard and pointer events, DOM mutations, timing signals, media state, screenshots, and accessibility-tree snapshots---\toolname{} combines persistent in-browser instrumentation with per-step evidence capture, rather than relying on the single audit state typically inspected by tools such as axe-core, WAVE, Lighthouse, and Pa11y~\cite{web:deque:axe-core,web:webaim:wave,web:google:lighthouse,web:pa11y:pa11y}. For judgment, prior LLM-based accessibility work shows that model reasoning can extend automated WCAG coverage~\cite{journal:gena11y:2025,journal:llm-manual-wcag-criteria:2025}; \toolname{} uses an LLM, in our case Gemini, more narrowly, invoking it only after an evidence gate succeeds and requiring every \textsc{fail} or \textsc{warning} to cite resolvable evidence references. The naive browser-use baseline and evidence-layer ablation therefore isolate the two claims tested in RQ2: the value of scenario execution with a browser agent, and the added value of criterion-specific, evidence-gated runtime analysis.

\subsection{Evaluation Denominators}

\Cref{tab:denominators} summarises the evaluation scope.
The benchmark contains 19 real public-web scenarios and 45 dynamic
WCAG checks per scenario, yielding 855 attempted scenario-criterion rows.
The execution completed 17 of 19 scenarios; 2 scenarios encountered
browser-task execution failures and produced no analysable trace.
The primary evaluation covers 765 scored rows ($17 \times 45$).
The 90 rows from failed scenarios are retained as an attempted-denominator
quantity but excluded from all primary metrics.

\begin{table}[H]
\centering
\caption{Evaluation scope and scoring denominators for the 19-scenario benchmark.}
\label{tab:denominators}
\begin{tabular}{lr}
\toprule
Quantity & Value \\
\midrule
Real public-web scenarios & 19 \\
Dynamic WCAG criteria per scenario & 45 \\
Attempted scenario-criterion rows & 855 \\
Successfully completed scenarios & 17 \\
Primary scored rows & 765 \\
Excluded rows (task failure) & 90 \\
Task success rate & 89.5\% \\
\bottomrule
\end{tabular}
\end{table}

%% file: sections/results.tex
\section{Results}
\label{sec:results}

\subsection{RQ1: Effectiveness on Real Interaction Flows}
\label{sec:results:rq1}

\toolname{} matched the manual oracle on 349 of 765 scored rows,
yielding an exact five-way status accuracy of 45.6\%
(95\% CI: 43.0--48.2\%).
This metric is strict: a prediction is counted correct only when the
analyzer returns the same label as the oracle from among
\textsc{fail}, \textsc{pass}, \textsc{warning}, \textsc{not\_applicable},
and \textsc{incomplete}.

\Cref{tab:primary-metrics} reports the full primary metric set.
The system achieves 41.4\% fail precision on 29 fail predictions against
31 oracle failures, and 79.0\% pass precision on 105 pass predictions
against 182 oracle pass rows.
Recall is lower for both outcome classes: fail recall is 38.7\% and pass
recall is 45.6\%.
The incomplete rate is 39.7\%: the analyzer abstains from a hard judgment
on 304 of 765 rows when the runtime evidence is insufficient to meet the
evidence gate.

\begin{rqresultbox}
\textbf{RQ1 result.}
On 765 scored scenario-criterion rows, \toolname{} achieved 45.6\%
exact five-way status accuracy, detected 12 of 31 oracle failures, and
abstained on 39.7\% of rows when required runtime evidence was
insufficient.
\end{rqresultbox}

\begin{table}[H]
\centering
\caption{Primary performance metrics over 765 scored scenario--criterion rows.}
\label{tab:primary-metrics}
\begin{tabular}{lrrr}
\toprule
Metric & Numerator & Denominator & Value \\
\midrule
Exact status accuracy  & 349 & 765 & 45.6\% \\
Fail precision         & 12  & 29  & 41.4\% \\
Fail recall            & 12  & 31  & 38.7\% \\
Fail F1                & --  & --  & 40.0\% \\
Pass precision         & 83  & 105 & 79.0\% \\
Pass recall            & 83  & 182 & 45.6\% \\
Pass F1                & --  & --  & 57.8\% \\
Not-applicable accuracy & 252 & 517 & 48.7\% \\
Incomplete rate        & 304 & 765 & 39.7\% \\
\bottomrule
\end{tabular}
\end{table}

The status distribution explains the result.
The oracle contains 517 not-applicable rows, 182 pass rows,
35 warning rows, and 31 fail rows.
\toolname{} predicts 292 not-applicable, 105 pass, 35 warning, 29 fail,
and 304 incomplete rows.
A substantial portion of mismatches therefore reflects conservative
abstention rather than incorrect prediction: the analyzer declines to
issue a pass or fail when the runtime evidence it collected does not
meet the criterion's evidence requirement.

\begin{table}[H]
\centering
\caption{Oracle and predicted status distributions over 765 scored rows.}
\label{tab:status-distribution}
\begin{tabular}{lrrr}
\toprule
Status & Oracle & \toolname{} & Naive audit \\
\midrule
Fail           & 31  & 29  & 3   \\
Warning        & 35  & 35  & 5   \\
Pass           & 182 & 105 & 34  \\
Not applicable & 517 & 292 & 0   \\
Incomplete     & 0   & 304 & 723 \\
\bottomrule
\end{tabular}
\end{table}

\subsection{RQ2: Benefit over a Generic Browser-Agent Audit}
\label{sec:results:rq2}

The naive browser-use global audit matches the oracle on 25 of 765 rows
(3.3\%), compared with 349 of 765 (45.6\%) for \toolname{}.
The difference is sharpest for failure detection: \toolname{} identifies
12 of 31 oracle failures, while the naive audit identifies none.
The naive audit marks 723 of 765 rows as \textsc{incomplete},
confirming that a generic browser agent does not naturally produce the
criterion-level runtime evidence required for dynamic WCAG assessment.

\begin{rqresultbox}
\textbf{RQ2 result.}
Compared with the browser-use baseline, \toolname{} improved exact accuracy
by 42.3 percentage points (45.6\% versus 3.3\%), detected 12 of 31 oracle
failures versus none, and reduced the incomplete rate from 94.5\% to 39.7\%.
These results show the value of a structured analysis
over unconstrained browser-agent exploration.
\end{rqresultbox}

\begin{table}[H]
\centering
\caption{\toolname{} and naive browser-use audit results on the same 765 oracle rows.}
\label{tab:baseline-comparison}
\begin{tabular}{lrr}
\toprule
Metric & \toolname{} & Naive audit \\
\midrule
Exact matches       & 349 / 765 & 25 / 765  \\
Exact accuracy      & 45.6\%    & 3.3\%     \\
Fail true positives & 12        & 0         \\
Fail predictions    & 29        & 3         \\
Oracle failures     & 31        & 31        \\
Fail precision      & 41.4\%    & 0.0\%     \\
Fail recall         & 38.7\%    & 0.0\%     \\
Incomplete rate     & 39.7\%    & 94.5\%    \\
\bottomrule
\end{tabular}
\end{table}

This result confirms the central claim of RQ2: the structured criterion
evidence pipeline provides measurable benefit over generic browser-use
exploration on the same oracle rows.
The useful unit of analysis is not a browser session but a
criterion-specific evidence packet grounded in a controlled interaction
trace.

\subsection{RQ3: Criterion-Family Capability and Failure Profile}
\label{sec:results:rq3}

\Cref{tab:family-results} reports exact-match accuracy and fail detection
by criterion family across the 765 scored rows.
Performance varies substantially by probe class.

\begin{table}[H]
\centering
\caption{Family-level outcomes on the 765 scored rows. 
Exact is the number of rows where \toolname{} matches the oracle label; 
Fail TP/Exp. reports correctly predicted failures over oracle failures.}
\label{tab:family-results}
\scriptsize
\setlength{\tabcolsep}{3pt}
\begin{tabularx}{\columnwidth}{Xrrrrr}
\toprule
Family & Crit. & Rows & Exact & Acc. & Fail TP/Exp. \\
\midrule
Media                  & 10 & 170 & 170 & 100.0\% & 0/0  \\
Keyboard/focus         & 10 & 170 & 100 & 58.8\%  & 10/23 \\
Pointer/gesture        & 5  & 85  & 47  & 55.3\%  & 0/0  \\
Status messages        & 1  & 17  & 8   & 47.1\%  & 0/2  \\
Hover/focus content    & 1  & 17  & 7   & 41.2\%  & 2/3  \\
Forms/errors/auth.     & 8  & 136 & 17  & 12.5\%  & 0/0  \\
Animation/flashing     & 3  & 51  & 0   & 0.0\%   & 0/0  \\
Timing/interruption    & 4  & 68  & 0   & 0.0\%   & 0/0  \\
Consistency/navigation & 3  & 51  & 0   & 0.0\%   & 0/3  \\
\bottomrule
\end{tabularx}
\end{table}
The keyboard and focus family produces the strongest results among
families with non-trivial oracle failure content: 100 exact matches
(58.8\%) and 10 true fail detections.
\Cref{tab:fail-by-criterion} details the within-family breakdown.
Focus order (SC~2.4.3), focus visible (SC~2.4.7), and focus appearance
(SC~2.4.13) each contribute confirmed-fail detections.
Hover and focus-revealed content (SC~1.4.13) also produces 2 of 3
oracle fail detections.

The media family achieves perfect accuracy, but this reflects the oracle
distribution rather than analytical strength: all scored media rows in
this benchmark are \textsc{not\_applicable}, and the evidence gate
correctly suppresses model calls for those criteria.
This result should not be interpreted as evidence of media-analysis
capability.

The zero-accuracy families -- animation, timing, and consistency -- expose
a systematic limitation of the current scenario set.
These criteria require targeted interactions or sustained observation
windows that the 19 scenarios did not consistently produce.
The analyzer applied the evidence gate correctly, returning
\textsc{incomplete} rather than making unsupported judgments;
but the gate itself cannot recover information that the trace does not
contain.

\begin{rqresultbox}
\textbf{RQ3 result.}
The system is strongest on keyboard/focus and hover/focus-revealed
content, where runtime interaction evidence directly supports judgment.
It currently struggles with animation, timing, and consistency criteria.
\end{rqresultbox}

\begin{table}[H]
\centering
\caption{Fail detection by individual criterion for criteria with oracle
failures or fail predictions.}
\label{tab:fail-by-criterion}
\begin{tabular}{lrrr}
\toprule
Criterion & TP & Predicted fail & Oracle fail \\
\midrule
1.4.13 Content on Hover or Focus & 2 & 3  & 3  \\
2.4.3  Focus Order               & 4 & 5  & 9  \\
2.4.7  Focus Visible             & 3 & 5  & 5  \\
2.4.13 Focus Appearance          & 3 & 4  & 4  \\
2.1.1  Keyboard                  & 0 & 4  & 1  \\
2.1.2  No Keyboard Trap          & 0 & 1  & 0  \\
2.1.3  Keyboard (No Exception)   & 0 & 1  & 3  \\
3.2.1  On Focus                  & 0 & 2  & 1  \\
3.3.1  Error Identification      & 0 & 3  & 0  \\
3.2.3  Consistent Navigation     & 0 & 0  & 1  \\
3.2.4  Consistent Identification & 0 & 0  & 2  \\
4.1.3  Status Messages           & 0 & 0  & 2  \\
\bottomrule
\end{tabular}
\end{table}

%% file: sections/discussion.tex
\section{Discussion}
\label{sec:discussion}
\subsection{Interpretation of Results}
\label{sec:discussion:interpretation}
The results show flow-aware accessibility assessment is both necessary
and achievable.
The naive browser-use audit -- operating without scenario specifications
or criterion-specific evidence collection -- detected none of the 31
oracle failures and left 94.5\% of rows \textsc{incomplete}, whereas
\toolname{} detected 12 of those failures and cut the incomplete rate
to 39.7\%: executing an interaction flow is necessary but not
sufficient, and criterion-specific evidence extraction paired with
grounded judgment closes the gap.

A practical deployment surfaces evidence-backed findings to prioritize
expert review rather than replace it.
Capability is strongest exactly where the design intends it: criteria
with direct behavioral evidence.
Keyboard and focus reaches 58.8\% exact-match accuracy and correctly
identifies 10 of 23 expected failures across focus order, visibility,
and appearance; hover and focus-revealed content identifies 2 of 3.
Performance is weaker for animation, timing, and consistency criteria,
which require sustained observation or repeated navigation the current
benchmark was not designed to reliably produce -- a concrete next step
rather than an open limitation. The system abstains here rather than
fabricating judgments, which is itself evidence the gating design works
correctly.

\subsection{Historical Comparison with the Pre-Layer Pipeline}
\label{sec:benchmark:ablation}

We compare the final pipeline with a prior run that used weaker evidence
structuring and did not include the criterion evidence packet,
fail-confidence validator, or calibrated status router.
The two runs differ in their completed scenarios and scored denominators
(630 versus 765 rows). Consequently, this comparison is descriptive and
does not isolate the causal effect of individual pipeline components.
\Cref{tab:ablation} summarises the observed differences.

\begin{table}[H]
\centering
\caption{Descriptive historical comparison of the pre-layer and final pipelines.}
\label{tab:ablation}
\begin{tabular}{lcc}
\toprule
Metric & Pre-layer run & Final pipeline \\
\midrule
Scored rows            & 630    & 765  \\
Exact status accuracy  & 57.5\% & 45.6\% \\
Fail precision         & 23.5\% & 41.4\% \\
Fail recall            & 72.7\% & 38.7\% \\
API Calls              & 609    & 203 \\
Invalid evidence refs  & 35     & 0     \\
\bottomrule
\end{tabular}
\end{table}

Relative to the pre-layer run, the final pipeline exhibits higher fail
precision (41.4\% versus 23.5\%), no invalid evidence references, and
fewer model calls. These changes coincide with lower fail recall and
exact status accuracy. Because the runs use different scored
denominators and differ in several pipeline components simultaneously,
these differences cannot be attributed exclusively to the evidence
layer.

Nevertheless, the comparison demonstrates the intended calibration
shift. The final pipeline reports fewer failures and abstains more
frequently, but its accepted findings cite resolvable runtime evidence.

For accessibility reports consumed by developers, a grounded confirmed
failure is more actionable than an unverified detection, justifying the
precision-over-recall tradeoff introduced by the evidence layer.

\subsection{Status Confusion Analysis}
\label{sec:results:confusion}

\Cref{tab:confusion} shows the full status confusion matrix.
The dominant error patterns are conservative: the analyzer frequently
returns \textsc{incomplete} or \textsc{not\_applicable} for rows the
oracle labels as \textsc{pass} or \textsc{fail}.
For 31 oracle-fail rows, the analyzer correctly returns \textsc{fail}
for 12, \textsc{warning} for 6, \textsc{incomplete} for 6, and
\textsc{pass} for 7.
The 7 pass predictions on oracle-fail rows represent true failures not
caught by the evidence gate or confirmed by the model.

\begin{table}[H]
\centering
\caption{Status confusion matrix. Rows are oracle labels; columns are
\toolname{} outputs. Primary scored rows: 765.}
\label{tab:confusion}
\begin{tabular}{lrrrrr}
\toprule
Oracle \textbackslash{} Output
  & Fail & Warn. & Pass & N/A & Incomp. \\
\midrule
Fail           & 12 & 6  & 7   & 0   & 6   \\
Warning        & 2  & 2  & 2   & 5   & 24  \\
Pass           & 12 & 15 & 83  & 35  & 37  \\
Not applicable & 3  & 12 & 13  & 252 & 237 \\
\bottomrule
\end{tabular}
\end{table}

\subsection{Model-Call Validity and Cost}
\label{sec:results:cost}

The dynamic WCAG analyzer made 203 Gemini calls over the final pipeline run.
The evidence-reference verifier accepted 201 calls and rejected 2
responses due to invalid JSON.
No model output contained an unresolvable evidence reference after \toolname{} verification, down from 35 invalid references in the pre-layer run.
The run consumed 17.1M prompt tokens and 31.1K output tokens for the
formal judge, with a total formal-judge runtime of approximately
2,746 seconds.

The total API cost for the full execution was approximately \texteuro{}8--12, corresponding to roughly \texteuro{}0.5 per completed scenario.

\begin{table}[H]
\centering
\caption{Validity and resource usage of 203 Gemini criterion judgments.}
\label{tab:model-cost}
\begin{tabular}{lr}
\toprule
Quantity & Value \\
\midrule
Formal model calls             & 203          \\
Valid verified calls            & 201          \\
Invalid JSON responses          & 2            \\
Unresolvable evidence refs      & 0            \\
Prompt tokens                   & 17,102,567   \\
Output tokens                   & 31,083       \\
Total tokens                    & 17,507,120   \\
Total formal-judge time (s)     & 2,746.16s     \\
Average per-call time (s)       & 13.53s         \\
\bottomrule
\end{tabular}
\end{table}

%% file: sections/threats.tex
\section{Threats to Validity}
\label{sec:threats}
% \Florian{separate section}
\subsection*{Construct validity}
The oracle consists of author judgments over runtime artifacts and
live website behavior, with reviewer identity, confidence, and
rationale recorded per entry. Interpretation remains possible for
criteria whose conformance depends on context outside the trace, and
the study does not execute screen readers or other assistive-technology
combinations.

\subsection*{Internal validity}
Each scenario was executed once; network conditions, website changes,
agent actions, and Gemini responses can alter the trace and final
classification across runs. Evidence-reference verification constrains
but does not eliminate this nondeterminism, a known property of
LLM-based judgment components.

\subsection*{External validity}

As they were not benchmark evaluating dynamic accessibility issues, we resorted to write one. The 19 curated scenarios cover diverse interaction patterns
but are not a random sample of websites and do not represent every
domain, browser, viewport, or assistive technology. However, they do represent concrete use-case for accessibility testing that were manually assessed by the authors. The results characterize \toolname{} on this
benchmark, an initial step toward broader evaluation rather than a
population-level estimate.

\subsection*{Conclusion validity}
Only 31 of 765 scored rows are oracle failures, so fail precision and
recall carry meaningful uncertainty and should be read alongside their
confidence intervals. Per-family results, based on smaller samples
still, are best read as observed capability boundaries rather than
stable estimates.

%% file: sections/artifact.tex
% \section{Artifact and Reproducibility}
% \label{sec:artifact}

% An anonymized artifact repository is available at \url{https://anonymous.4open.science/r/flow-a11y/} containing the \toolname{} source code, benchmark fixtures, oracle labels, prompt templates, an executable notebook, and normalized metrics tables that reproduce the reported results without browser interaction or model calls. Full end-to-end reruns require a Gemini API key, Chrome-compatible browser environment, browser-use support, and access to live websites; regenerated traces may therefore differ from the saved evaluation.

%% file: sections/conclusion.tex
\section{Conclusion}
\label{sec:conclusion}

This paper presented \toolname{}, a flow-aware accessibility
testing system for WCAG criteria whose assessment depends on user
interaction.
By treating an interaction scenario, rather than a loaded page, as
the unit of analysis, the system collects multimodal runtime traces,
gates model invocation on evidence sufficiency and applicability,
verifies that model findings cite resolvable evidence, and exports
auditable criterion-level artifacts.

On 19 real public-web scenarios and 765 scored oracle rows,
\toolname{} achieves 45.6\% exact status accuracy and detects 12 of
31 oracle failures at 41.4\% precision, substantially outperforming
a naive browser-use audit (3.3\% accuracy, no detected failures).
The evidence-calibration layer trades recall for precision and
eliminates invalid evidence references entirely -- a deliberate
choice for developer-facing reports, where unverified findings carry
real triage cost.

\toolname{} complements, rather than replaces, page-level scanners:
together, static and flow-aware analysis cover more of the WCAG
criterion space than either alone.
Future work should target the zero-accuracy criterion families
through criterion-aware scenario design.